\documentclass[aps,pre,one,11pt,showpacs]{revtex4}
\usepackage{epsfig}
\usepackage{graphicx}

\begin{document}

\title{Generalized hydrodynamics of a dilute finite-sized particles
suspension: Dynamic viscosity}
\author{S. I. Hern\'{a}ndez$^{1}\thanks{%
Corresponding author: Phone: 52 (+55) 56 22 47 23; Fax: 52 (+55) 56 16 12
01; E-mail: ivan@iim.unam.mx}$, I. Santamar\'{\i}a-Holek$^{2}$, Carlos I.
Mendoza$^{1}$ and L. F. del Castillo$^{1}$}
\affiliation{$^{1}$Instituto de Investigaciones en Materiales, Universidad Nacional Aut%
\'{o}noma de M\'{e}xico, Apdo. Postal 70-360, 04510 M\'{e}xico, D.F., Mexico}
\affiliation{$^{2}$Facultad de Ciencias, Universidad Nacional Aut\'{o}noma de M\'{e}xico,
Circuito Exterior de Ciudad Universitaria, 04510, M\'{e}xico D. F., Mexico}

\begin{abstract}
We present a mesoscopic hydrodynamic description of the dynamics of
colloidal suspensions. We consider the system as a gas of Brownian particles
suspended in a Newtonian heat bath subjected to stationary non-equilibrium
conditions imposed by a velocity field. Using results already obtained in
previous studies in the field by means of a generalized Fokker-Planck
equation, we obtain a set of coupled differential equations for the local
diffusion current and the evolution of the total stress tensor. We find that
the dynamic shear viscosity of the system contains contributions arising
from the finite size of the particles.
\end{abstract}

\maketitle


\section{Introduction}

\noindent Due to the theoretical and industrial importance of colloidal
suspensions, the rheological properties of these systems have been the
subject of intense investigations by using a variety of experimental and
theoretical methods \cite{bird,ottingerlibro,libro-edwards-beris,jou2,powell}%
. Among others, the experimental techniques used to study these systems are
neutron and light scattering, and ultrasonic absorption \cite%
{vanderwerff,cichocki1,cichocki2}. From the theoretical point of view,
non-equilibrium statistical mechanics and thermodynamics \cite%
{lutsko,bedeaux2,bird,ISHPRE,CPI,libro-edwards-beris,ottingerlibro,jou2},
mode-coupling formalism \cite{miyazaki} and computer simulations \cite%
{ladd,groot-warren,boek,barrat,todd} are current techniques in which
particular interest has been put on the behavior of the frequency-dependent
shear viscosity of semidilute suspensions of hard spheres in different
situations involving direct or hydrodynamic interactions \cite%
{cichocki1,cichocki2}. 

In this article, we present a mesoscopic hydrodynamic description of the
dynamics of colloidal suspensions subjected to stationary non-equilibrium
conditions imposed by a velocity field. In particular, we analyze the
behavior of the dynamic viscosity due to its importance in reflecting the
microstructure details into the relaxation of the system under the
conditions imposed by the applied flow. At mesoscopic level, the
contributions of the suspended phase to the dynamics of the whole system can
be described by using a Fokker-Planck equation \cite%
{ISHPRE,PNAS,rosalio,oppenheim}. This equation has been previously obtained
on the grounds of irreversible thermodynamics in the dilute regime \cite%
{ISHPRE,PNAS}. However, it is important to stress that similar equations
have been derived in the context of different approximations ranging from
the kinetic theory of gases to the projector operator formalism \cite%
{rosalio,oppenheim,drossinos}. From this Fokker-Planck equation we calculate
a set of hydrodynamic equations for the moments of the distribution
function. Then, by using the fluctuating hydrodynamics approach we obtain a
frequency dependent correction to the dynamic viscosity of the system that
accounts for memory effects \cite{mazur,boon}.

The paper is organized as follows.\ In Section II we summarize the
derivation of the Fokker-Planck equation and carry out the hydrodynamic
description of a gas of Brownian particles. In Section III we derive the
dynamic viscosity of the system and present numerical results. Finally,
Section IV is devoted to conclusions.

\section{Kinetic and hydrodynamic description of a gas of Brownian particles}

We will analyze the dynamics of a system consisting in a \textquotedblleft
gas\textquotedblright\ of Brownian particles of mass $m$ suspended in a
newtonian heat bath subjected to stationary non-equilibrium conditions
imposed by a stationary flow $\vec{v}_0(\vec{r})$.

The usual starting point to describe the dynamics of this system is the $N-$%
particle distribution function and its corresponding multivariate
Fokker-Planck equation. In this equation, microscopic expressions for the
hydrodynamic and direct interactions must be introduced. In the semidilute
regime, for example, one must introduce the Oseen tensor and the
two-particle interaction potential. At higher concentrations, the subsequent
approximations may, in general, consider the Rotne-Prager tensor and the
three-particle (or higher) interaction potential. It is clear that this
approach constitutes a $N-$body problem hard to solve by analytical methods 
\cite{lionberger,brady2000}.

Here, we will follow an alternative approach to the problem in terms of a
mean field approximation in which the difficulties of the $N-$body problem
are avoided. This approximation, based on the one-particle distribution
function, allows one to make some analitycal progress when calculating the
hydrodynamic equations, despite that the explicit dependence of the
transport coefficients on frequency and wave vector and the form of the
direct interactions are \emph{a priori} unknown. However, by assuming in
first approximation that the transport coefficients appearing in the
hydrodynamic equations are constants, we may use of the well established
formalism of fluctuating hydrodynamics to obtain the explicit expressions
for the corrections to the transport coefficients as a functions of the
frequency and the wave vector \cite{mazur,boon,ISHJPCM}.

According to this, at mesoscopic level the dynamics of the dilute Brownian
gas can be described in terms of a single-particle local probability
distribution $P(\vec{r},\vec{u},t)$ depending on time $t$ and the
instantaneous position and velocity of the Brownian particle $\vec{r}$ and $%
\vec{u}$. This distribution function obeys the continuity equation%
\begin{equation}
\frac{\partial }{\partial t}P+\nabla \cdot \left( \vec{u}P\right) =-\frac{%
\partial }{\partial \vec{u}}\cdot \left( P\vec{V}_{\vec{u}}\right) ,
\label{contphasespace}
\end{equation}%
where $\vec{V}_{\vec{u}}$ is a streaming velocity in $\vec{u}-$space. The
explicit form of $\vec{V}_{\vec{u}}$ can be obtained from the entropy
production of the system, after assuming valid the hypothesis of local
equilibrium in phase-space and using the rules of mesoscopic non-equilibrium
thermodynamics \cite{PNAS,SteadyState}. Then, entropy production can be
calculated from the nonequilibrium entropy functional \cite{ISHJPCM}%
\begin{equation}
\rho s(\vec{r},t)=-k_{B}\int P(\vec{u},\vec{r},t)\ln \frac{z}{z_{le}}d\vec{u}%
+\rho s_{le}(\vec{r}),  \label{p. gibbs}
\end{equation}%
where $k_{B}$ is Boltzmann's constant and $\rho s(\vec{r},t)$ and $\rho
s_{le}(\vec{r})$ are the nonequilibrium entropy per unit volume and the
local equilibrium entropy per unit volume, respectively. Moreover, we have
introduced the fugacities $z=P(\vec{r},\vec{u},t)A(\rho )$ and $z_{le}=\exp
\left\{ [\mu _{le}-\frac{1}{2}\left( \vec{u}-\vec{v}_{0}\right)
^{2}]m/k_{B}T\right\} $ with $\mu _{le}$ the chemical potential at a local
equilibrium and $\rho (\vec{r},t)=m\int P(\vec{u},\vec{r},t)d\vec{u}$ the
mass density field of the Brownian gas. Finally, $A(\rho )$, related with
the excess of osmotic pressure of the system, represents the activity
coefficient that introduces the interactions among Brownian particles \cite%
{ISHJPCM}.

After calculating the entropy production using (\ref{p. gibbs}), identifying
forces and currents and using the linear-law scheme of non-equilibrium
thermodynamics \cite{degroot}, one finds that the distribution function $P(%
\vec{r},\vec{u},t)$ obeys the Fokker-Planck equation 
\begin{equation}
\frac{\partial }{\partial t}P+\nabla \cdot \left( \vec{u}P\right) -\nabla %
\left[ \frac{k_{B}T}{m}\ln A(\rho )\right] \cdot \frac{\partial P}{\partial 
\vec{u}}=\frac{\partial }{\partial \vec{u}}\cdot \left[ \vec{\vec{\beta}}%
\cdot (\vec{u}-\vec{v}_{0})P+\frac{k_{B}T}{m}\vec{\vec{\xi}}\cdot \frac{%
\partial P}{\partial \vec{u}}\right] ,  \label{F-P}
\end{equation}%
where $k_{B}T$ is the thermal energy (assumed constant). In the general
case, the friction tensor $\vec{\vec{\beta}}(\vec{r},t)$ must contain the
average effect of hydrodynamic interactions. It is related to the Onsager
coefficient $\vec{\vec{\xi}}(\vec{r},t)$ through $\vec{\vec{\beta}}=\vec{%
\vec{\xi}}+\vec{\vec{\epsilon}}\cdot \nabla \vec{v}_{0}$, in which the
Onsager coefficient $\vec{\vec{\epsilon}}(\vec{r},t)$ accounts for the
effects of the force exerted by a fluid on a particle of finite size
according to the Fax\'{e}n theorem \cite{ISHPRE,SteadyState}.

It is important to notice that $\vec{v}_{0}$ is the stationary solution of
the Navier-Stokes equation $\rho _{0}\vec{v}_{0}\nabla \vec{v}_{0}=-\nabla
\cdot \!\!\,\,\mathbf{P}^{0}\!\!\,\,$ for the heat bath, with the
appropriate boundary conditions on the surface of the colloidal particle and 
$\rho _{0}$ is the constant density of the bath.

Equations similar to (\ref{F-P}) have been derived in the context of
different approximations, see for example, Refs. \cite%
{ISHPRE,SteadyState,ISHJPCM,rosalio,oppenheim}. From equation (\ref{F-P}),
one may calculate a set of evolution equations for the moments of the
probability distribution, giving the \textquotedblleft
hydrodynamic\textquotedblright\ description of the system \cite%
{ISHPRE,ISHJPCM}. After taking into account the usual definitions for the
hydrodynamic-like fields: mass density $\rho (\vec{r},t)$, velocity $\vec{v}(%
\vec{r},t)$ and pressure tensor $\!\!\,\mathbf{P}^{k}(\vec{r},t)$ (see
reference \cite{ISHPRE}), the first equation of this set corresponds to the
conservation of mass 
\begin{equation}
\frac{\partial \rho }{\partial t}=-\nabla \cdot \vec{J}_{D},
\label{contmasa}
\end{equation}%
where the diffusion current $\vec{J}_{D}(\vec{r},t)=\rho \vec{v}$ of the
Brownian gas satisfies the evolution equation \cite{ISHJPCM,SteadyState} 
\begin{equation}
\vec{J}_{D}+\vec{\vec{\beta}}^{-1}\cdot \frac{D\vec{J}_{D}}{Dt}=\rho \vec{v}%
_{0}-\frac{k_{B}T}{m}\vec{\vec{\beta}}^{-1}\cdot \frac{\partial \ln A(\rho )%
}{\partial \ln \rho }\nabla \rho -\vec{\vec{\beta}}^{-1}\cdot \left[ \nabla
\cdot \,\,\!\!\mathbf{P}^{k}\!\!\,\,(\vec{r},t)\right] .  \label{diffusive}
\end{equation}%
Here $D/Dt$ is the material time derivative, $k_{B}$ is Boltzman's constant,
and $T$ is the temperature of the heat bath. The matrix of relaxation times
of the diffusion current is: $\vec{\vec{\tau}}_{1}\simeq \vec{\vec{\beta}}%
^{-1}$. Finally, Eq. (\ref{diffusive}) contains the pressure tensor $\!\!\,%
\mathbf{P}^{k}\!\!(\vec{r},t)$ of the suspended phase whose evolution
equation is \cite{ISHPRE} 
\begin{equation}
\left[ \!\!\,\,\mathbf{P}^{k}\!\!\,\,\cdot \vec{\vec{\tau}}_{2}^{-1}\right]
^{s}=\frac{k_{B}T}{m}\rho \left[ \vec{\vec{\beta}}-\vec{\vec{\epsilon}}\cdot
\nabla \vec{v}_{0}\right] ^{s}-\frac{1}{2}\frac{D}{Dt}\!\!\,\,\mathbf{P}%
^{k}\!\!\,\,,  \label{evolution-P}
\end{equation}%
where the symbol $s$ stands for the symmetric part of a tensor and $\vec{%
\vec{\tau}}_{2}$ is a matrix of relaxation times given by 
\begin{equation}
\vec{\vec{\tau}}_{2}\simeq \left( \vec{\vec{\beta}}+\nabla \vec{v}\right)
^{-1},  \label{tau1}
\end{equation}%
where we have neglected inertial terms proportional to $\nabla \cdot \vec{v}$%
, assumption valid for small density fluctuations $\Delta \rho /\rho \ll 1$.

The explicit expression for the pressure tensor can be obtained by solving
the set of differential equations (\ref{evolution-P}). However, in order to
make analytical progress, let us assume the approximation $\vec{\vec{\tau }}%
_{2}\simeq \vec{\vec{\beta}} ^{-1} \cdot \left( \mathbf{1}-\vec{\vec{\beta}}
^{-1} \cdot \nabla \vec{v}\right) $ with $\mathbf{1}$ the unit tensor, and
which allows to write Eq. (\ref{evolution-P}) in the form

\begin{equation}
\!\!\,\,\mathbf{P}^{k}\!\!\,\, \simeq \frac{k_{B}T}{m}\rho \mathbf{1}%
-\left\{ \vec{\vec{D}}_{0} \rho \cdot \left[ \nabla \vec{v}+\left( \vec{\vec{%
\epsilon}} \cdot \nabla \vec{v}_{0}\right)^{\dag}\right]\right\} ^{s}-\frac{1%
}{2} \left\{\left(\vec{\vec{\beta}}^{-1}\right)^{\dag} \cdot \frac{D }{Dt}
\!\!\,\,\mathbf{P}^{k}\!\!\,\,\right\} ^{s},  \label{evol-Parrelgada2}
\end{equation}%
where the $\dag$ represents the transpose, we have defined the diffusion
coefficient $\vec{\vec{D}}_{0}=\frac{k_{B}T}{m}\vec{\vec{\beta}}^{-1}$ and
assumed that the leading term in the matrix of relaxation times is $\vec{%
\vec{\beta}}^{-1}$ . Now, the assumption of small density fluctuations
implies that the diagonal terms of $\!\!\,\,\mathbf{P}^{k}\,\,$ relax faster
than the non-diagonal ones. Therefore, Eq. (\ref{evol-Parrelgada2}) can be
separated in the form $\!\!\,\mathbf{P}^{k}\,=p^{id}\mathbf{1}+\mathbf{L}$,
with $\mathbf{L}(\vec{r},t)$ the traceless stress tensor. This separation
leads to the equations 
\begin{equation}
p^{id}=\frac{k_{B}T_{B}}{m}\rho \,\,\,\,\,\,\,\,\,\,\,\,{\text{and}}%
\,\,\,\,\,\,\,\,\,\,\,\,\!\!\,\,2 \mathbf{L}\!\!\,\,+ \left[ \left(\vec{\vec{%
\beta}}^{-1}\right)^{\dag} \cdot \frac{D}{Dt}\!\!\,\,\mathbf{L}\,\,\right]
^{s}=- 2 \left\{\vec{\vec{D}}_{0}\rho \cdot \left[ \nabla \vec{v}%
\,^{s}+\left( \vec{\vec{\epsilon}} \cdot \nabla \vec{v}_{0}\right)^{\dag}%
\right]\right\} ^{s} .  \label{evol-L}
\end{equation}%
After substitution of equations (\ref{evol-L}) into (\ref{diffusive}) we
obtain 
\begin{equation}
\vec{J}_{D}=\rho \vec{v}_{0}-\vec{\vec{D}}(\rho )\cdot\nabla \rho -\vec{\vec{%
\beta}}^{-1}\cdot \left( \nabla \cdot \mathbf{L}\,\,\right)-\vec{\vec{\beta}}%
^{-1}\cdot\frac{D\vec{J}_{D}}{Dt},  \label{diffusive15}
\end{equation}%
where we have defined the collective diffusion coefficient \cite{vanderwerff}
\begin{equation}
\vec{\vec{D}}(\rho )=\vec{\vec{D}}_{0} \left[ 1+\frac{\partial \ln A(\rho )}{%
\partial \ln \rho }\right] =\vec{\vec{D}}_{0}\frac{\partial \mu (\rho )}{%
\partial \rho },  \label{Dcol}
\end{equation}%
in accordance with Ref. \cite{degroot}. Notice that in the limit of very
dilute suspensions, the activity coefficient vanishes and then Eq. (\ref%
{diffusive15}) recovers the expected form.

The evolution equations governing the behavior of the hydrodynamic fields
describing the whole suspension as an effective medium can be obtained from
Eqs. (\ref{evol-L}) and (\ref{diffusive15}) by adding the local diffusion
current $\vec{J}=\vec{J}_{D}-\rho \vec{v}_{0}$ and the traceless part of the
stress tensor of the heat bath given by $\!\!\,\,\mathbf{P}%
^{0}\,\,\!\!=-2\eta _{0}\left( \nabla \vec{v}_{0}\right) ^{0}$ with $\eta
_{0}$ the Newtonian viscosity of the bath \cite{bedeaux}.

Thus, multiplying Eq. (\ref{evol-L}) by $\rho _{0}\rho ^{-1}$, adding $%
\!\!\,\,\mathbf{P}^{0}\,\,\!\!$, neglecting the convective term in the total
time derivative and using the definition of $\vec{J}$, one arrives at the
evolution equation for the total traceless stress tensor of the suspension $%
\!\!\,\,\mathbf{Q}^{0}\!\!\,\,\equiv \!\!\,\mathbf{\,P}^{0}\!\!\,\,+2\!\!\,%
\,\rho _{0}\rho ^{-1}\mathbf{L}\!\!\,\,$ 
\begin{equation}
\!\!\,\,\mathbf{Q}^{0}\,\,\!\!+\left[ \left( \vec{\vec{\beta}}^{-1}\right)
^{\dag }\cdot \frac{\partial }{\partial t}\!\!\,\,\mathbf{Q}\,\,\!\!\right]
^{0}=-2\left[ \vec{\vec{D}}_{0}\rho _{0}\rho _{B}^{-1}\cdot \left( \nabla 
\vec{J}\right) \right] ^{0}-2\left[ \vec{\vec{\eta}}_{fs}\cdot (\nabla \vec{v%
}_{0})\,\right] ^{0},  \label{evol-Q}
\end{equation}%
where $\rho _{B}=mn$, being $n=N/V$ the number density of colloidal
particles, and we have defined the effective viscosity coefficient $\vec{%
\vec{\eta}}_{fs}$ of the whole suspension as 
\begin{equation}
\vec{\vec{\eta}}_{fs}\mathbf{=\,}\eta _{0}\left[ \mathbf{1}+\vec{\vec{D}}%
_{0}\rho _{0}\eta _{0}^{-1}\cdot \left( \mathbf{1}+\vec{\vec{\epsilon}}%
\right) \right] .  \label{visceffect}
\end{equation}%
Equation (\ref{visceffect}) contains corrections due to the finite size of
the particles through the coefficient $\vec{\vec{\epsilon}}\simeq \left(
m/6k_{B}T\right) a^{2}\vec{\vec{\beta}}\cdot \vec{\vec{\beta}}$ with $a$ the
radius of the particle. The expression for $\vec{\vec{\epsilon}}$ has been
obtained by taking into account the Fax\'{e}n theorem \cite{SteadyState}. In
deriving Eqs. (\ref{evol-Q}) and (\ref{visceffect}) we have assumed that
terms of the order $\vec{\vec{D}}_{0}\vec{J}\rho ^{-2}\cdot \nabla \rho $
may be neglected and that the coefficient of the first term at the
right-hand side of (\ref{evol-Q}) only involves the average constant value
of the density of the suspended phase. This is consistent with the
approximation $\Delta \rho /\rho \ll 1$ used to derive Eq. (\ref{evolution-P}%
).

Now, by substituting $\vec{J}_{D}=\vec{J}+\rho \vec{v}_{0}\,$ into Eq. (\ref%
{diffusive15}) and using the Navier-Stokes equation, one can derive the
evolution equation for the local diffusion current $\vec{J}$ 
\begin{equation}
\left[ \vec{\vec{\beta}}+\nabla \vec{v}_{0}\right] \cdot \vec{J}+\frac{D\vec{%
J}}{Dt}=-\vec{\vec{\beta}}\cdot \left( \vec{\vec{D}}_{eff}\cdot \nabla \rho
\right) +\rho \rho _{0}^{-1}\nabla \cdot \!\!\,\,\mathbf{Q}_{qs}^{0}\!\!\,\,.
\label{diffusive2}
\end{equation}%
where we have introduced the effective diffusion tensor 
\begin{equation}
\vec{\vec{D}}_{eff}=\vec{\vec{D}}(\rho )-\vec{\vec{\beta}}^{-1}\cdot \vec{%
\vec{D}}_{0}\cdot \left\{ \left[ (\mathbf{1}+\vec{\vec{\epsilon}})\cdot
\nabla \vec{v}_{0}\right] ^{0}+\rho ^{-1}(\nabla \vec{J})^{0}\right\} ,
\label{difusion effectiva}
\end{equation}%
and $\!\!\,\,\mathbf{Q}_{qs}^{0}\!\!\,\,$ is the quasi-stationary form of
the stress tensor of the suspension whose explicit form follows from Eq.(\ref%
{evol-Q}) for times $t\gg (\beta ^{-1})_{ij}$. The dependence of $\vec{\vec{D%
}}_{eff}$ on the stress tensor of the heat bath has been previously reported
in Refs. \cite{rosalio,ISHPRE,drossinos}. Now, in accordance with the
assumption of low velocity gradients, the term $\vec{\vec{\beta}}^{-1}\cdot
\nabla \vec{v}_{0}$ can be neglected in Eqs. (\ref{diffusive2}) and (\ref%
{difusion effectiva}), and then we obtain the simplified equation%
\begin{equation}
\vec{J}+\vec{\vec{\beta}}^{-1}\cdot \frac{D\vec{J}}{Dt}\simeq -\vec{\vec{D}}%
(\rho )\cdot \nabla \rho +\rho _{B}\rho _{0}^{-1}\vec{\vec{\beta}}^{-1}\cdot
\left( \nabla \cdot \!\!\,\,\mathbf{Q}^{0}\!\!\,\,\right) ,
\label{diffusive3}
\end{equation}%
where we have assumed that $\!\!\,\,\mathbf{Q}^{0}\,\,\!\!$ can be
introduced instead of $\!\!\,\,\mathbf{Q}_{qs}^{0}\!\!\,\,$, and that the
coefficient of the last term only involves the average constant value of the
density of the suspended phase $\rho _{B}$.

Equations similar to (\ref{evol-Q}) and (\ref{diffusive3}) have been
discussed by several authors from different macroscopic and microscopic
points of view in the context of polymer solutions \cite%
{beris1,beris2,ottinger,bhave}. These equations incorporate viscoelastic
effects into the description since they contain a term associated with the
relaxation of the involved quantities ($\!\!\,\,\mathbf{Q}^{0}\,\,\!\!$, $%
\vec{J}$), and thus lead to more general constitutive relations than those
for Newtonian fluids. Equations (\ref{evol-Q}) and (\ref{diffusive3})
include tensor coefficients being then suitable to describe anisotropic
systems \cite{jou1,jou2}. In the following section, we will use equations (%
\ref{evol-Q}) and (\ref{diffusive3}) to analyze the rheological properties
of the system when fluctuations of the suspended phase are considered.

\section{The Dynamic Viscosity}

The dynamic viscosity coefficient of the suspension as a function of the
frequency $\omega $, can be calculated by considering the linearized
fluctuating equations for the hydrodynamic fields which follow from
equations (\ref{contmasa}), (\ref{evol-Q}) and (\ref{diffusive3}). With this
purpose, we will assume that $\rho =\rho _{B}+\delta \rho $, $\vec{v}_{0}=%
\vec{v}_{s}+\delta \vec{u}$ and $\mathbf{Q}^{0}=\mathbf{Q}_{0}^{0}+\delta 
\mathbf{Q}^{0}$, with $\delta \rho $, $\delta \vec{u}$ and $\delta \mathbf{Q}%
^{0}$ the deviations with respect to the average values $\rho _{B}$, $\vec{v}%
_{s}$ and $\mathbf{Q}_{0}^{0}$.

Moreover, in this section we will consider the case of a dilute suspension
of particles. This assumption leads to a simplification of the description
in which both hydrodynamic and direct interactions can be neglected. As a
consequence, the set of Onsager and transport coefficients become scalar
and, in the simplest approximation, constant: $\vec{\vec{\beta}}=\beta 
\mathbf{1}$, $\vec{\vec{\epsilon}}=\epsilon \mathbf{1}$, $\vec{\vec{\xi}}%
=\xi \mathbf{1}$ and $\vec{\vec{\eta}}_{fs}=\eta _{fs}\mathbf{1}$, and the
activity coefficient $A(\rho )=1$, thus simplifying the expression for the
diffusion coefficient to $D(\rho )=D_{0}$.

Then, at first order in the deviations, from Eqs. (\ref{contmasa}), (\ref%
{evol-Q}) and (\ref{diffusive3}) one obtains the following set of
fluctuating hydrodynamic equations 
\begin{equation}
\frac{\partial \left( \delta \rho \right) }{\partial t}=-\nabla \cdot \left(
\delta \vec{J}\right) -\nabla \rho _{0}\cdot \delta \vec{u}-\nabla \left(
\delta \rho \right) \cdot \vec{v}_{s},  \label{deltarho}
\end{equation}

\begin{equation}
\delta \vec{J}+\beta ^{-1}\frac{D}{Dt}\delta \vec{J}=-D_0 \nabla \delta \rho
+\beta ^{-1} \rho_B \rho_0^{-1} \nabla \cdot \delta \mathbf{Q}^{0},
\label{deltaj}
\end{equation}

\begin{equation}
\delta \mathbf{Q}^{0}+\beta ^{-1}\frac{D}{Dt}\delta \mathbf{Q}%
^{0}=-2D_{0}\rho _{B}\rho _{0}^{-1}\left( \nabla \delta \vec{J}\,\right)
^{0}-2\mathbf{\eta }_{fs}\left( \nabla \delta \vec{u}\,\right) ^{0}.
\label{deltaq}
\end{equation}

To obtain the corrections to the dynamic viscosity of the suspension, Eq. (%
\ref{deltaq}) can be written in a more suitable form by taking into account
the separation\textbf{\ $\vec{v}_{0}=\vec{v}_{s}+\delta \vec{u}$}, which can
be interpreted as the sum of the unperturbed imposed velocity field\textbf{\ 
$\vec{v}_{s}$ }, in particular, a shear flow, and\textbf{\ $\delta \vec{u}$ }%
the corresponding correction to the velocity field arising from the presence
of the particle, which is proportional to\textbf{\ $\nabla \vec{v}_{s}$ }%
\cite{landau}. Following Ref. \cite{landau}, Eq. (\ref{deltaq}) can be
written in the form 
\begin{equation}
\delta \mathbf{Q}^{0}+\beta ^{-1}\frac{D}{Dt}\delta \mathbf{Q}%
^{0}=-2D_{0}\rho _{B}\rho _{0}^{-1}\left( \nabla \delta \vec{J}\,\right)
^{0}-5\phi \mathbf{\eta }_{fs}\left( \nabla \vec{v}_{s}\,\right) ^{0}.
\label{deltaq2}
\end{equation}%
where $\phi $ represents the volume fraction occupied by the Brownian
particles in the whole system.

Now, by taking the Fourier transform of Eqs. (\ref{deltarho}), (\ref{deltaj}%
) and (\ref{deltaq2}) and solving for $\delta \mathbf{Q}^{0}$ in terms of $i%
\vec{k}\delta \vec{u}$, and taking into account that the frequency-dependent
correction to the viscosity $\Delta \eta \left( \vec{k},\omega \right) $ of
the suspension is defined according to the relation 
\begin{equation}
\delta \mathbf{Q}^{0}=-i[\Delta \mathbf{\eta }\left( \vec{k},\omega \right) 
\vec{k}\delta \vec{u}]^{0},  \label{dynamicviscosity1}
\end{equation}%
we find that 
\begin{equation}
\Delta \eta \left( \vec{k},\omega \right) \simeq \frac{5}{2}\phi \frac{%
\mathbf{\eta }_{fs}}{1-i\omega \beta ^{-1}-\frac{D_{0}\beta ^{-1}k^{2}}{%
\left( 1-i\omega \beta ^{-1}\right) }},  \label{dynamicviscosity2}
\end{equation}%
where, for simplicity, we have considered the case in which $\vec{k}=\left(
k,0,0\right) $. Notice that the frequency dependence of the dynamic
viscosity for semidiluted suspensions follows a second order
continued-fraction expansion \cite{cichocki2}. If we now take into account
the relation $\mathbf{Q}^{0}=\mathbf{Q}_{0}^{0}+\delta \mathbf{Q}^{0}$, we
obtain that the total viscosity of the suspension\textbf{\ $\eta
_{fs}+\Delta \eta \left( k,\omega \right) $ }can be written as 
\begin{equation}
\mathbf{\eta (}k,\mathbf{\omega )}\simeq \eta _{fs}\left[ 1+\frac{5}{2}\phi 
\frac{1}{1-i\omega \beta ^{-1}-\frac{D_{0}\beta ^{-1}k^{2}}{\left( 1-i\omega
\beta ^{-1}\right) }}\right] .  \label{viscfinal}
\end{equation}%
It is interesting to notice that by expanding Eq. (\ref{viscfinal}) up to
first order in $\omega $ and second order in $k$ and taking into account Eq.(%
\ref{visceffect}), we obtain 
\begin{equation}
\mathbf{\eta (}k,\mathbf{\omega })\mathbf{\simeq \,}\eta _{0}\left[ 1+\frac{5%
}{2}\phi \left( 1+\frac{i\omega }{\beta }+k^{2}\frac{D_{0}}{\beta }\right) %
\right] \left[ 1+\tilde{Sc}^{-1}\left( 1+\epsilon \right) \right] ,
\label{viscexpan}
\end{equation}%
where $\tilde{Sc}\equiv \eta _{0}/(\rho _{0}D_{0})$ is the Schmidt number
for the Brownian gas diffusing in the solvent. Eq. (\ref{viscexpan}) is an
expression similar to that reported in Ref. \cite{bedeaux}. Nonetheless, Eq.
(\ref{viscexpan}) includes corrections due to the finite size of the
particles which are not included in the theories of Refs. \cite{vanderwerff}
and \cite{bedeaux}.

To analyze the frequency behavior of Eq. (\ref{viscfinal}), it can be
recasted in a more convenient form by using the dispersion relation $k^{2}=-i%
\frac{\omega \rho _{0}}{\eta _{0}}$, which can be obtained from the
Navier-Stokes equations \cite{landau,SteadyState}. Substituting the last
expression into Eq. (\ref{viscfinal}) and using dimensionless variables
defined by 
\begin{equation}
\widetilde{\omega }\equiv \frac{\omega }{\beta }\,,\,\,\,\,\,\,\,\,%
\widetilde{k}^{2}\equiv \frac{D_{0}}{\beta }k^{2}\,\,\,\,\,\,{\text{and}}%
\,\,\,\,\,\,\widetilde{\eta }(\widetilde{k},\widetilde{\omega })\equiv \frac{%
\eta (k,\omega )}{\mathbf{\eta }_{fs}},  \label{variables}
\end{equation}%
we obtain the following expression for the dynamic viscosity 
\begin{equation}
\widetilde{\eta }\left( \widetilde{\omega }\right) \simeq 1+\frac{5}{2}\phi 
\frac{1}{1-i\widetilde{\omega }+i\frac{\widetilde{\omega }}{\tilde{Sc}\left(
1-i\widetilde{\omega }\right) }},  \label{viscomega}
\end{equation}%
which only depends on the dimensionless frequency $\widetilde{\omega }$, the
volume fraction $\phi $, and the Schmidt number $\tilde{Sc}$. The term
containing $\tilde{Sc}$ represents a contribution arising from the coupling
between diffusion and viscosity. Note that our definition of the Schmidt
number is related to that used in Ref. \cite{arxiv} by means of the relation 
$\tilde{Sc}=ScD_{f}/D_{0}$, being $D_{f}$ the fluid self-diffusion
coefficient. For a gas $Sc\approx 1$ (this value is also used in numerical
simulations of colloidal systems as, for example, the one of Ref. \cite%
{groot-warren}), whereas for a liquid $Sc\gg 1$ and since for a colloidal
suspension in general $D_{f}\gg D_{0}$ then $\tilde{Sc}\gg 1$. Therefore,
the terms containing $\tilde{Sc}^{-1}$ in Eqs.(\ref{visceffect}) and (\ref%
{viscomega}) can be neglected and the expression for the dynamic viscosity
reduces to a form similar to that of Ref. \cite{vanderwerff}, the
differences being the finite size contributions contained in $\mathbf{\eta }%
_{fs}$, the explicit dependence on the volume fraction and that the
characteristic relaxation time is given by $\beta ^{-1}$ instead of $\tau
_{D}=a^{2}/D_{0}$.

In Fig. \textbf{1} we show the real $\widetilde{\eta }^{\prime }\left( 
\widetilde{\omega }\right) $ and imaginary $\widetilde{\eta }^{\prime \prime
}\left( \widetilde{\omega }\right) $ parts of the complex dynamic viscosity $%
\widetilde{\eta }\left( \widetilde{\omega }\right) $ as a function of $\log 
\widetilde{\omega }$ for different values of the volume fraction and
considering $\tilde{Sc}\gg 1$. Notice that $\widetilde{\eta }\left( 
\widetilde{\omega }\right) $ has the usual behavior of a relaxation
processes with only one relaxation time \cite{vanderwerff}. From Eq.(\ref%
{viscomega}) it can be seen that the ratio $\left[ \widetilde{\eta }^{\prime
}\left( \widetilde{\omega }\right) -1\right] /\left[ \widetilde{\eta }%
_{0}^{\prime }-1\right] $ , with $\widetilde{\eta }_{0}^{\prime }\equiv 
\widetilde{\eta }^{\prime }\left( \widetilde{\omega }=0\right) $, will
depend only on the dimensionless parameter $\widetilde{\omega }$. The
imaginary term can be reduced in the same manner $\widetilde{\eta }^{\prime
\prime }\left( \widetilde{\omega }\right) /\left[ \widetilde{\eta }%
_{0}^{\prime }-1\right] $. In Fig. \textbf{2} we plot the reduced viscosity
as a function of $\widetilde{\omega }$. In addition to the usual case of
very large Schmidt numbers (solid line), we have also explored what would be
the consequences when the Schmidt number is of the order of unity. For $%
\tilde{Sc}=1$ (dashed line) the real part of the viscosity presents a hump
near $\widetilde{\omega }=1$. This characteristic appears for values of $%
\tilde{Sc}$ in the range $[1,2+\sqrt{3}]$ whereas for $\tilde{Sc}\geq 2+%
\sqrt{3}$ the hump is not present (for $\tilde{Sc}\geq 1$, the imaginary
part of $\widetilde{\eta }\left( \widetilde{\omega }\right) $ takes only
positive values). At intermediate frequencies $\widetilde{\eta }^{\prime
}\left( \widetilde{\omega }\right) $ decays following the law $\widetilde{%
\omega }^{-2}$, independently of the value of $\tilde{Sc}$. This is a
typical behavior for processes with only one relaxation time \cite%
{vanderwerff}. On the other hand, the imaginary part of the viscosity has a
bell-shaped form for $\tilde{Sc}\gg 1$, whereas for $\tilde{Sc}=1$ it
becomes asymmetric and more stretched. It is important to notice that for
intermediate frequencies, $\widetilde{\eta }^{\prime \prime }\left( 
\widetilde{\omega }\right) $ decays as $\widetilde{\omega }^{-1}$, also a
usual behavior for processes with only one relaxation time. The low
frequency region of the dynamic viscosity curves deserves also a detailed
analysis. In this frequency region, $\widetilde{\eta }^{\prime \prime
}\left( \widetilde{\omega }\right) $ grows as $\widetilde{\omega }$,
independently of the value of the Schmidt number, except for $\tilde{Sc}=1$
where the curve grows as $\widetilde{\omega }^{3}$.

Finally, it is important to point out that it has been experimentally
observed that for intermediate frequencies both, the real and imaginary
parts of the viscosity decay as $\omega ^{-1/2}$ \cite{vanderwerff}. Such a
frequency behavior can be modeled by assuming a distribution of relaxation
times $\tau _{p}.$ In this case, the viscosity can be written as 
\begin{equation}
\widetilde{\eta }\left( \omega \right) =1+\sum\limits_{p=1}^{N}\frac{%
G_{p}\tau _{p}}{1-i\omega \tau _{p}+i\frac{\omega \tau _{p}}{\tilde{Sc}%
\left( 1-i\omega \tau _{p}\right) }},
\end{equation}%
where $N$ is the number of relaxation processes involved. $G_{p}$ is a
constant relaxation strength of the relaxation process with relaxation time $%
\tau _{p}$, given by a distribution of the form 
\begin{equation}
\tau _{p}=\tau _{1}p^{-\alpha },\text{ \ \ \ \ \ }\alpha =2
\end{equation}%
with $\tau _{1}$ the longest relaxation time \cite{vanderwerff}. For a
non-interacting hard-sphere system this would correspond to a distribution
on the size of the particles $a_{p}=a_{1}p^{\alpha }$, or equivalently, $%
\tilde{Sc}_{p}=\tilde{Sc}_{1}p^{\alpha }$. With this size distribution our
results for the viscosity lead to the same power law decay on frequency $%
\omega ^{-1/2}$. In Fig. \textbf{3} we show this behavior for two different
values of the Schmidt number corresponding to the longest relaxation time $%
\tau _{1}$, $\tilde{Sc}_{1}=1,$ and $\tilde{Sc}_{1}\gg 1$. We notice that
the intermediate frequency power-law dependency on frequency is not altered
by this choice but at low frequencies differences shows up for $\widetilde{%
\eta }^{\prime \prime }\left( \omega \right) $. A different power law for
the size distribution will modify the behavior of the viscosity.

\section{Conclusions}

We have proposed a model for the dynamics of colloidal suspensions based on
a mesoscopic hydrodynamic description derived by applying the rules of
mesoscopic non-equilibrium thermodynamics. We have derived the viscosity of
a monodisperse hard-sphere system that includes effects due to the finite
size of the colloidal particles. It has been shown that the normalized
dynamic viscosity depends only on three dimensionless parameters: the
Schmidt number $\tilde{Sc}$, the volume fraction $\phi $, and the normalized
frequency $\widetilde{\omega }$. Also, it presents power-law exponents for
different regions of the frequency space. Our results show that the
mesoscopic approach leads to results which are consistent with previous
theoretical descriptions derived by using the generalized hydrodynamics
theory. We expect that the present model may be extended to a wider class of
suspensions including particles with internal structure.

\qquad {\LARGE Acknowledgements}

This work was supported in part by Grants DGAPA IN-119606 (LFC and SIH),
IN-110103 (CIM) and IN-108006 (ISH). Also by the CONACYT under Grants
SEP-2004-C01-47070 (LFC and SIH) and 43596-F (CIM).

\newpage

\begin{figure}[tbp]
{}
\par
\centering\mbox{\resizebox*{7cm}{!}{\includegraphics{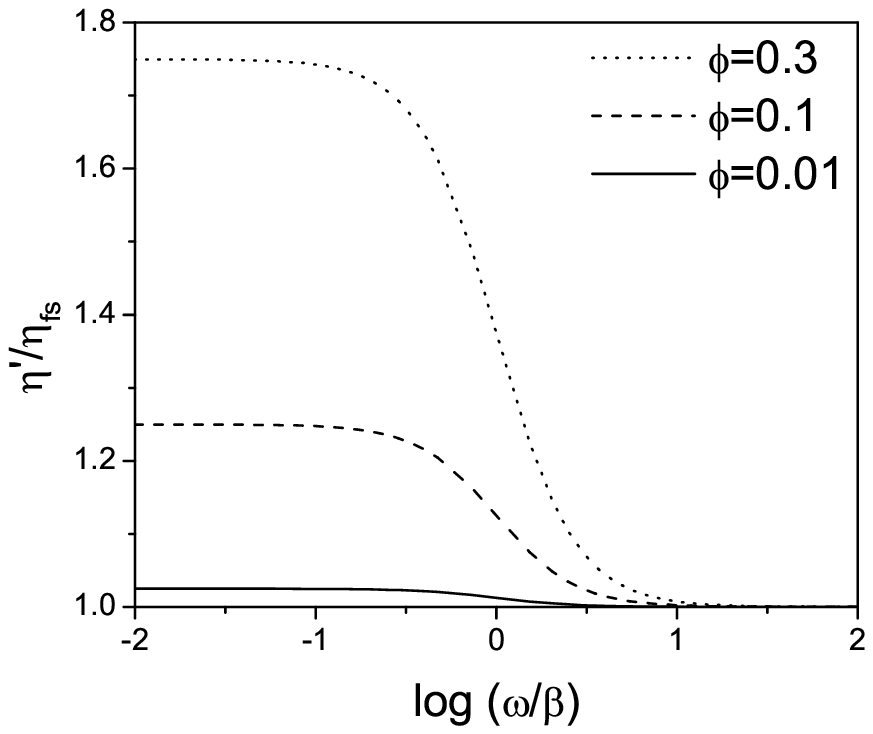}} } %
\mbox{\resizebox*{7cm}{!}{\includegraphics{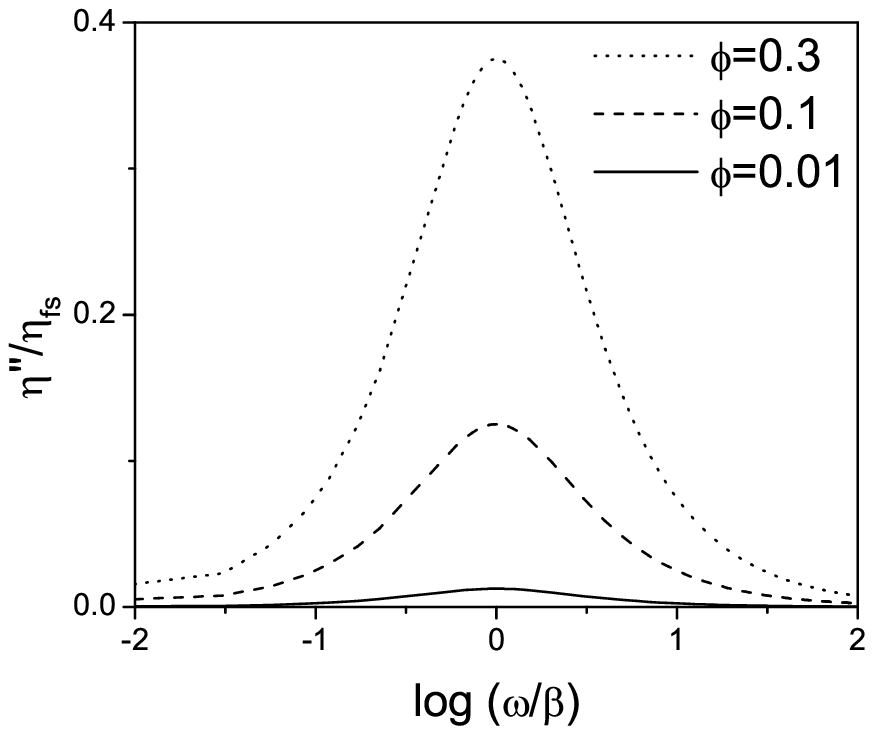}} }
\par
{\footnotesize {\ } \vspace{0.0cm} }
\caption{(a) Real and (b) imaginary parts of the normalized dynamic
viscosity as a function of $\widetilde{\protect\omega }$ for $\tilde{Sc}\gg
1 $. Volume fractions as indicated in the figures.}
\label{fig1}
\end{figure}

\begin{figure}[tbp]
{}
\par
\centering\mbox{\resizebox*{7.0cm}{!}{\includegraphics{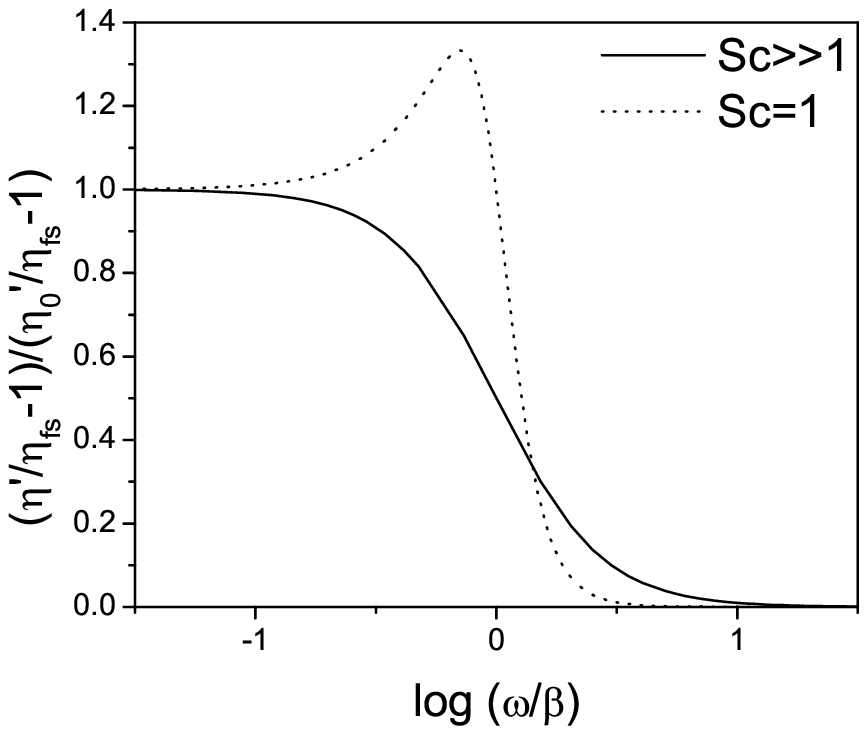}} } %
\mbox{\resizebox*{7.0cm}{!}{\includegraphics{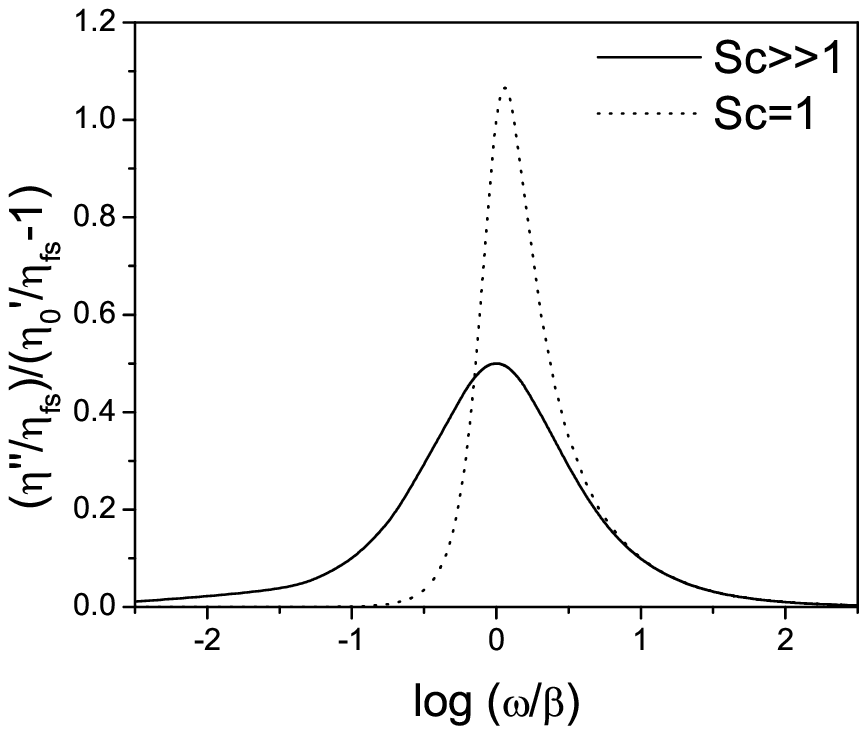}} }
\par
{\footnotesize {\ } \vspace{0.0cm} }
\caption{(a) Real and (b) imaginary parts of the reduced dynamic viscosity
as a function of $\widetilde{\protect\omega }$. The solid line corresponds
to $\tilde{Sc}\gg 1$ and the dashed line corresponds to $\tilde{Sc}=1$.}
\label{fig2}
\end{figure}

\begin{figure}[tbp]
{}
\par
\centering\mbox{\resizebox*{7.0cm}{!}{\includegraphics{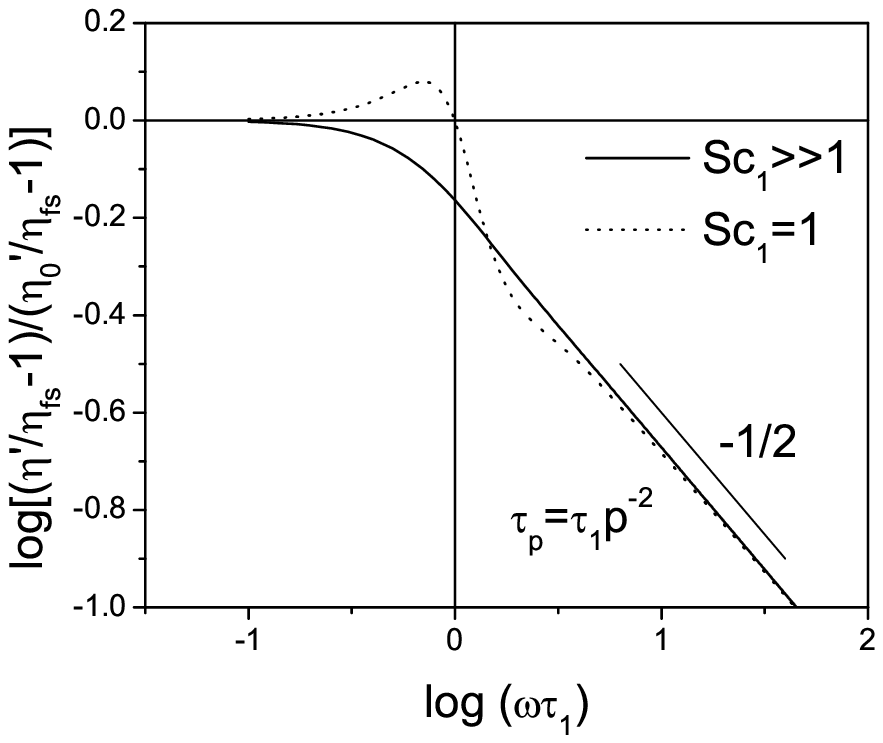}} } %
\mbox{\resizebox*{7.0cm}{!}{\includegraphics{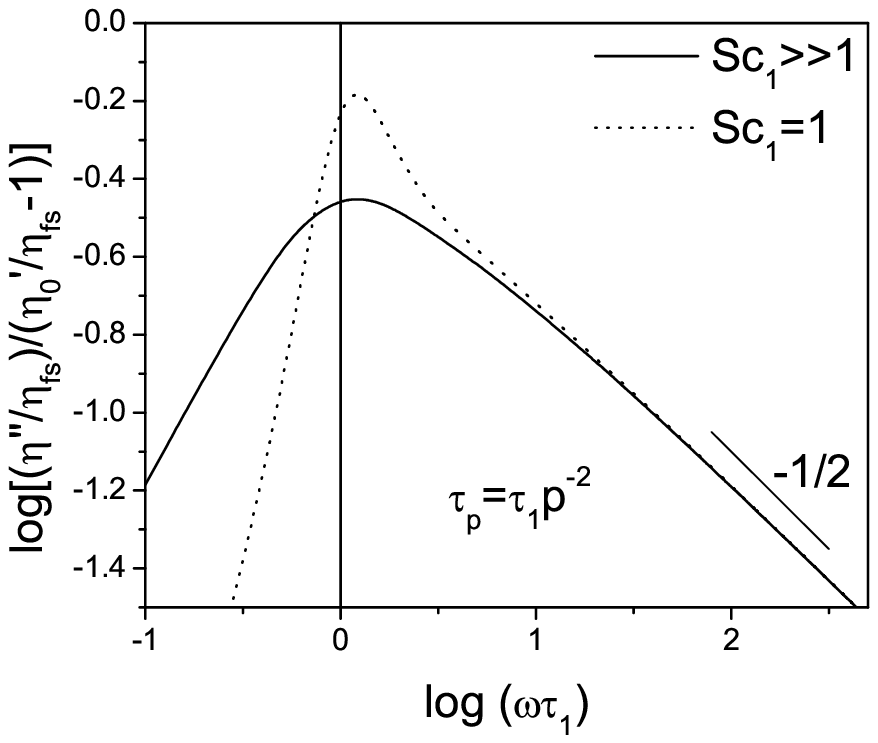}} }
\par
{\footnotesize {\ } \vspace{0.0cm} }
\caption{(a) Real and (b) imaginary parts of the reduced dynamic viscosity
as a function of $\widetilde{\protect\omega }$, for a distribution of
relaxing processes. The solid line corresponds to $\tilde{Sc}_{1}\gg 1$ and
the dashed line corresponds to $\tilde{Sc}_{1}=1$. The number of relaxing
processes considered was $N=1000$.}
\label{fig3}
\end{figure}


\begin{references}

\bibitem{bird} R.B. Bird, O. Hassager, {\it Dynamics of Polymeric Liquids} 
(Wiley-Interscience, New York, 1987).

\bibitem{ottingerlibro} H.C. Ottinger {\it Stochastic Processes in Polymeric Fluids} 
(Springer, New York, 1996).

\bibitem{libro-edwards-beris} A.N. Beris and B.J. Edwards {\it Thermodynamics of Flowing Systems} (Oxford University Press, New York, 1994).

\bibitem{jou2} D. Jou, J. Casas Vazquez and G. Lebon, {\it Extended
Irreversible Thermodynamics}, (Springer, Berlin, 1996). 

\bibitem{powell} J. J. Stickel and R. L. Powell, Annu. Rev. Fluid Mech. {\bf 37}, 129 (2005)

\bibitem{cichocki1} B. Cichocki, B. U. Felderhof, Phys Rev. A {\bf 43}, 5405
(1991).

\bibitem{cichocki2} B. Cichocki, B. U. Felderhof, Phys Rev. A {\bf 46}, 7723
(1992).

\bibitem{vanderwerff} J.C. van der Werff et. al., Phys. Rev. A {\bf 39}, 795
(1989).

\bibitem{lutsko} J.F. Lutsko and J.W. Dufty, Phys. Rev. E {\bf 66}, 041206
(2002); M.H. Ernst, Phys. Rev. E {\bf 71}, 030101 (2005); P. Das and J.K.
Bhattacharjee, Phys. Rev. E {\bf 71}, 036145 (2005).

\bibitem{ISHPRE} I. Santamar\'{\i}a-Holek, D. Reguera and J. M. Rub\'{\i},
Phys. Rev. E {\bf 63} 051106 (2001).

\bibitem{CPI} C. M\'alaga, F. Mandujano and I. Santamar\'{\i}a-Holek,
Physica A {\bf 369}, 291 (2006).

\bibitem{bedeaux2} D. Bedeaux, J.M. Rubi, Physica A {\bf 305}, 360 (2002)
and references therein.

\bibitem{miyazaki} K. Miyazaki, et. al., Phys. Rev. E {\bf 70}, 011501
(2004).

\bibitem{ladd} A.J.C. Ladd, Phys. Rev. Lett. {\bf 70}, 1339 (1993).

\bibitem{groot-warren} R.D. Groot and P.B. Warren, J. Chem. Phys. {\bf 107}, 4423 (1997).

\bibitem{boek} E.S. Boek, et. al., Phys. Rev. E {\bf 55}, 3124 (1997).

\bibitem{barrat} J.-L. Barrat and L. Berthier, Phys. Rev. E {\bf 63}, 012503
(2000); J. Chem. Phys. {\bf 116}, 6228 (2002).

\bibitem{todd} B.D. Todd, Phys. Rev. E {\bf 72}, 041204 (2005).

\bibitem{PNAS} D. Reguera, J. M. Rub\'{\i} and J. M. G. Vilar, J. Phys. Chem. B {\bf %
109}, 21502 (2005).

\bibitem{rosalio} R.F. Rodr\'{\i}guez, E. Sal\'{\i}nas-Rodr\'{\i}guez, and J.W. Dufty, 
J. Stat. Phys. {\bf 32}, 279 (1983).

\bibitem{oppenheim} J.E. Shea and I. Oppenheim, Physica A {\bf 250}, 265 (1998).

\bibitem{drossinos} Y. Drossinos, M. W. Reeks, Phys. Rev. E {\bf 71}, 031113 (2005).

\bibitem{mazur} P. Mazur and D. Bedeaux, Physica {\bf 76}, 235 (1974). 

\bibitem {boon} J. P. Boon, S. Yip,{\it Molecular Hydrodynamics}, (Dover, New York, 1991).

\bibitem {lionberger} R. A. Lionberger, W. B. Russel, Adv. Chem. Phys. {\bf 111}, 399 (2000).

\bibitem {brady2000} J. F. Brady, J. Rheol. {\bf 44}, 629 (2000).

\bibitem{ISHJPCM} J. M. Rub\'{\i}, I. Santamar\'{\i}a-Holek and A. P\'{e}%
rez-Madrid , J. Phys.-Condes. Matter {\bf 16}, S2047 (2004).

\bibitem{SteadyState} I. Santamar\'{\i}a-Holek, J. M. Rub\'{\i} and A. P\'{e}%
rez-Madrid, New J. Phys. {\bf 7}, 35 (2005).

\bibitem{degroot} S. R. de Groot, P. Mazur, {\it Non-equilibrium
Thermodynamics} (Dover, New York, 1984).

\bibitem{landau} L. D. Landau and E. M. Lifshitz, \emph{Course of Theoretical Physics, Fluid Mechanics} (Pergammon, New York 1980), Vol. 6.

\bibitem{beris1} A. N. Beris and B. G. Mabratzas, J. Rheol. {\bf 38}, 1235
(1994).

\bibitem{beris2} A. N. Beris and B. J. Edwards, {\it Thermodynamics of
Flowing Systems}, (Oxford University Press, Oxford, 1991).

\bibitem{ottinger} H. C. Ottinger, Rheol. Acta {\bf 31}, 14 (1992).

\bibitem{bhave} A. V.Bhave, R.C. Armstrong and R. A. Brown, J. Chem. Phys. 
{\bf 95}, 2988 (1991).

\bibitem{jou1} D. Jou, J. Camacho and M. Grmela, Macromolecules {\bf 24},
3597 (1991).


\bibitem{bedeaux} D. Bedeaux, R. Kapral and P. Mazur, Physica A {\bf 88}, 88
(1977).

\bibitem{arxiv} J.T. Padding and A.A. Louis, arXiv:cond-mat/0603391.

\end{references}
\end{document}